\input harvmac
\centerline{\bf Beating the Standard Model$^*$}\medskip
\centerline{\bf Sheldon L. Glashow}
\centerline{\bf Harvard University}

\footnote{}{$^*$ This research was supported in part by the National
Science Foundation under grant NSF-PHYS-92-18167.}

{\bigskip\narrower\noindent\sl  This report, adapted from my talk at the
1998 Ettore Majorana Subnuclear School at Erice, proffers speculative
explanations of the strong CP problem and the existence of cosmic rays
beyond the GZK bound. It is based on works done with
Sidney Coleman and Howard Georgi.\medskip}

\newsec{Introduction} Although our beloved standard model 
of quarks and leptons  offers a complete, consistent and
correct description of most particle phenomena, lots of 
vexing questions remain unanswered, such as:  Why are
the gauge group and the  fermion masses what they are? Why three
families of quarks and leptons?  What breaks electroweak symmetry? 
How about gravity? Leaving  such profound
meta-questions to supersymmetrists,
string theorists and their successors, we shall beat
on two unrelated and more modest puzzles: 
the origin of CP violation and
the reported observation of 
cosmic-ray events with unexpectedly high energies.\bigskip

\newsec{The Strong CP Problem} 
The standard model  admits two kinds of CP violation: (1)
Complex  Yukawa couplings of the Higgs doublet  produce CP-violating mass
matrices for up-like and down-like quarks and generate the complex phase
$\delta$ appearing in the Kobayashi-Maskawa (KM) matrix. This mechanism
offers a plausible explanation of all observed CP-violating effects. \ (2)
The CP-odd self interaction of the QCD gauge fields characterized by the
$\theta$ parameter (together with complex quark masses) can induce 
additional CP
violating effects such as a neutron electric dipole moment. Although the
complex phase $\delta$ is large in the standard model, this
parameter must be unnaturally small, to wit: 
$\bar\theta\equiv \theta +{\rm Arg\,det}\,
M < 3\times 10^{-10}\,,$ where $M$
is the quark mass matrix. 

The strong CP problem is often addressed,
but many proposed solutions seem to me  either
contrived, inelegant, empirically unacceptable, or all three. 
Several recent papers approach the problem in a novel
manner~\ref\rgg{H.  Georgi and S.L. Glashow, {\tt hep-ph/9807399} and work
in progress.} \ref\rbc{See also {\it e.g.,}  P.H. Frampton and M. Harada,
{\tt hep-ph/9809402}\semi D. Bowser-Chau,  D. Chang and W.-Y. Keung, Phys.
Rev. Lett. 81 (1998) 2028.}.  I  shall discuss one 
model of this kind developed 
with Howard Georgi.
The key notion is for CP to be softly broken: 
conserved by all terms in the Lagrangian with mass dimension
four, but not by terms with  lower mass dimension.  Of course, the only
lower dimension term in the standard-model Lagrangian is the Higgs mass,
which cannot violate CP. Our hypothesis excludes both 
sources of symmetry violation;
$\bar\theta=0$ and the KM matrix is real.
We've lost the baby with
the bathwater; there is no strong CP problem because  CP is unbroken.

To do better, additional architecture  is needed:
new particles with new interactions,
but not too many of them!  The new particles are heavy because they
are not seen. They---unlike
quarks and leptons---have gauge invariant mass terms with
mass dimension less than four. Therein lurks the origin of
CP violation!

Wait a second! Heavy unobserved particles are often posited for
other reasons. The aesthetic appeal of
left-right symmetry, or of
$SO(10)$ over $SU(5)$, suggests that the 15-member families of chiral fields 
be extended to include massive singlet neutrinos with large
Majorana masses.  These heavy neutrinos 
generate see-saw Majorana masses for the three weak-doublet neutrinos, 
from the interplay between
the Higgs mass terms linking doublet and singlet neutrinos 
and much larger gauge-invariant Majorana mass terms linking singlet
neutrinos to one another.

The bare mass term of the left-handed singlet neutrinos  $N_i$ has
mass-dimension three and (according to our hypothesis) need not, indeed
should not, conserve  CP. Thus heavy neutrino masses are described by a
{\it complex\/}  $3\times 3$ symmetric matrix
${\cal M}_{ij}$. Ordinary neutrinos acquire the masses $m^\dagger{\cal
M}^{-1}m$, where $m$ is the unknown   Higgs-generated real matrix  with
eigenvalues sometimes, without much reason, taken comparable to charged
lepton masses.  A unitary matrix analogous to the  complex  KM matrix of
the quark sector in the standard model  relates neutrino mass and flavor
eigenstates.  Thus the see-saw mechanism leads directly to CP violation in
the realm of neutrino physics,  where with luck it may be
seen,   but hardly at all in  the  relevant realm of hadron physics.

To save ourselves,
the  seemingly useless  CP
violation buried in the heavy neutrino  mass matrix 
somehow must contaminate the
quark sector. Remarkably, just one more
unobserved particle does  the trick:
a heavy boson  $\zeta$  which is a color
triplet and a weak doublet with electric charges ${2\over3}$ and 
$-{1\over3}$. The $\zeta$ has Yukawa interactions with
coupling constants  $g_{ia}$ linking  each flavor $i$ of quark doublet
to each flavor $a$  of Majorana neutrino.

To make order-of-magnitude
estimates, we tentatively take 
the masses of the $\zeta$ meson and the three singlet neutrinos   
comparable
and $\sim\! M_\zeta$. 
We also take the $g_{ia}$ comparable
and $\sim \!g$. What we know about quark and lepton masses suggests that
this approximation is unsound: we use it only 
to get a feel for the model.

As in the standard model, the leading contribution to  CP violation among
kaons comes from a box diagram by which two $s$ quarks become two
$d$ quarks.  But it's not the usual box diagram  with internal quark and
$W$-boson lines because the KM matrix is real in tree
approximation, and as we shall see, remains 
nearly real when radiatively corrected.
Rather, it's the box with internal $N$'s and
$\zeta$'s. For this diagram to generate the observed value of 
$\epsilon_K$,  the common mass and coupling constant must satisfy~\rgg\
the constraint:
\eqn\enum{  {\alpha_g\over M_\zeta}\approx 2\times 10^{-8}\;\rm GeV\,,}
with $\alpha_g = g^2/4\pi$.

CP-violating radiative corrections to the renormalizable interactions of
the standard model must be small if our unconventional
 box diagram is to yield
all observable CP violation. In particular,
they induce finite complex phases in the KM  matrix. 
Some phases can be removed by field redefinitions, 
but the area $\cal A$ of the unitarity triangle,
given by
$$2\,{\cal A}= \bigg\vert {\rm
Im}\,\big(V_{ub}V_{ud}^*V_{cb}^*V_{cd}\big)\bigg\vert\,,$$
is an invariant measure of CP violation in the KM matrix. 
For models such as this one,
we have shown~\rgg\  $\cal A$ to be tiny compared to its value in the standard
model. The radiatively corrected
KM matrix is nearly real and the
unitarity triangle is an 
essentially straight line. Salvatore Mele finds
this result to be compatible with all 
available experimental data~\ref\rmele{Salvatore Mele, {\tt hep-ph/9808411}
and {\tt hep-ph/9810333}.}. 
If a model of this kind is correct, experimenters at BELLE and BABAR
are in for a big surprise.\medskip

What about strong CP? Although $\bar \theta=0$ in the bare
Lagrangian, radiative corrections $\Delta M$ to
the quark mass matrix can be complex.
Current constraints on the neutron electric dipole moment 
require:
$\Delta \bar\theta\equiv {\rm Arg\,det}\, M < 3\times 10^{-10}\,.$ 
Because corrections to quark masses are small, this condition becomes:
$${\rm Im}\, \bigg[{\rm tr}\,
\big(\Delta M_U^{\vphantom{-1}}M_U^{-1} +
\Delta M_D^{\vphantom{-1}}M_D^{-1}\big)\bigg]< 3\times 10^{-10}\,.$$
For the model at hand, we have shown~\rgg\ that the first non-vanishing
contribution to $\Delta \bar\theta$ appears at three loops. 
Estimating this diagram, we found that the 
strong CP
problem is  solved  provided that $\alpha_g 
< 0.0024$.

Making use of Eq.~\enum,  we find the bound  $ M_\zeta< 1.2\times 
10^5$~GeV, which is a bit awkward from the point of view of
see-saw neutrino masses. However, the present analysis is merely a proof of
principle that our model can solve the problem.
We can relax our assumption that all heavy-sector
masses are comparable. Surely there may be a hierarchy of
singlet neutrino masses, whereupon
the leading contribution to $\Delta \bar \theta$ is suppressed by an
additional ratio of singlet neutrino masses.
Much larger values of $\alpha_g$,
and much smaller see-saw neutrino masses, can be obtained without
encountering a strong CP problem. 
\bigskip

\newsec{Ultra-High Energy Cosmic Rays}

Primary nucleons with
sufficient energy will collide inelastically with CBR photons, thereby
losing energy. 
This  results in the GZK cutoff, saying that nucleons
with energies $>5\times10^{19}\;$eV cannot reach us from distances
greater that $\sim 50$~Mpc. However, cosmic rays are seen
well above this energy~\ref\rcr{AGASA 
Collab. {\it e.g.,} M. Takeda
{\it et.al.,} Phys. Rev. Lett. 81 (1998) 1163\semi
Fly's Eye Collab. {\it e.g.,} D.J. Bird {\it et al.,} Astrophys. J. 424 (1995
144\semi Haverah Park Collab. See: M.A. Lawrence {\it et al.,} J. Phys. G17
(1991) 733\semi Yakutsk Collab. {\it e.g.,} N.N. Efimov {\it et al.,} in
22nd Intl. Cosmic Ray Conf. (1991) Dublin.}.
Indeed, there are a handful of events with energies significantly
above $10^{20}$~eV. In this connection, a remarkable correlation has been
discovered by  Farrar and Biermann:
that the five highest energy cosmic ray events seem to be
closely aligned in space with
compact radio-loud quasars
at cosmological distances~\ref\rgf{G.R. Farrar and P.L. Biermann,Phys. Rev.
Lett. 81 (1998) 3579.}.

We argue that these events may have been produced by
ultra-high-energy (UHE) primary {\it neutrons\/} 
that are both stable and immune to
the GZK cutoff. To accomplish these miracles, we invoke 
tiny departures from strict Lorentz invariance, 
too small to
have been detected otherwise.  The results in this section are abstracted
from a recent paper with Sidney
Coleman~\ref\rcg{S. Coleman and S.L. Glashow, HUTP-98/A082
{\tt hep-ph/9812418.} See also the references therein.}. 
Many observable  consequences
of Lorentz violation are described in terms of  modified
energy-momentum relations for freely moving particles. To each particle
species $a$ there corresponds a mass $m_a$ and a maximum attainable velocity
$c_a$. (Here we neglect the possibility that $c_a$
may be helicity-dependent and flavor non-diagonal. Furthermore, we do not
consider violations of TCP symmetry.)
The dispersion
relations become $E^2= c_a^2\vec p^{\,2}+ m_a^2c_a^2$.  Lorentz invariance
is recaptured iff all $c_a$ are the same.

Ordinarily, free neutrons can beta decay but protons cannot. 
Departures from Lorentz invariance can affect the
kinematics of decay processes and even invert 
this pattern! To see how, let's examine the case
$c_p=c_e=c_\nu < c_n$, where conventional relativistic kinematics
may be used with
$c_p$ as ``the speed of light,'' provided the
neutron is assigned an effective mass $m_{\rm eff}\;$  given by:
$ m^2_{\rm eff} \equiv m_n^2-(c_p^2-c_n^2)\,\vec p^{\,2}\,,$
where $\vec p$ is its momentum in the preferred frame. 
Neutron beta-decay is allowed iff
$m_{\rm eff}> m_p+m_e$. Expressed in terms of
the neutron energy $E$ in the preferred frame, this condition
becomes:
$$ E<E_1=\sqrt{{m_n^2 -(m_p+m_e)^2\over c_p^2-c_n^2}}
\simeq 
2.7\times 10^{19}\; \left[{10^{-24}\over c_p-c_n}\right]^{1/2}\; \rm eV. $$
With our choice of Lorentz-violating parameters, \ 
{\it neutrons with energies exceeding $E_1$ are 
stable particles that can
be present among UHE cosmic rays.}

Conversely, we find that a proton with energy $E$ can
beta decay iff:
$$ E>E_2\simeq \sqrt{{m_n^2-(m_p-m_e)^2\over c_p^2-c_n^2}}
\simeq 
4.1\times 10^{19}\ \left[{10^{-24}\over c_p-c_n}\right]^{1/2}\ \rm eV. $$
For our example,
{\it protons with energies exceeding 
$E_2$ are unstable particles that cannot be present among UHE
cosmic rays.}
The above results are expressed in terms of
a nominal choice, $c_p-c_n= 10^{-24}$, lying
beyond the sensitivity of current tests of Lorentz invariance. Perhaps
highest energy cosmic-ray
primaries are stable neutrons.
\medskip

Next we point out that  there may not be a GZK cutoff. 
Effects of departures from Lorentz invariance 
increase rapidly with energy and can 
kinematically prevent cosmic-ray nucleons from undergoing inelastic
collisions with CBR photons. The cutoff thereby  undone,  a deeply
cosmological origin of UHE cosmic rays becomes  tenable.
To see how this goes, consider the formation reaction yielding
the first pion-nucleon resonance: 
\eqn\eform{ p+ \gamma\ {\rm(CBR)}\longrightarrow \Delta(1232)\,,} 
by a  proton of energy 
$E$ colliding with a CBR photon of energy $\omega$.  The target photons
are thermal with $T=2.73\;$K or
$kT\equiv\omega_0=2.35\times 10^{-4}\;{\rm eV}$. 
For a head-on impact, $\Delta$ formation is allowed iff: 
\eqn\eforma{2\omega +{M_p^2\over 2E}\ge (c_\Delta-c_p)\,E+{M_\Delta^2\over
2E}\,,}
where $c_\Delta-c_p$ is the relevant Lorentz-violating parameter.
If $c_\Delta=c_p$,  Eq.~\eforma\  yields the usual threshold,
$E_f=(M_\Delta^2-M_p^2)/4\omega$. Otherwise, 
Eq.~\eforma\ yields  a quadratic inequality in $E$ which can be satisfied
iff  $c_\Delta-c_p<\hat \delta(\omega)\equiv \omega/2E_f$.
As $c_\Delta-c_p$ increases toward $\hat \delta$,
the threshold for $\Delta$ formation  grows. If it 
exceeds its critical value,
\eqn\edelcrit{c_\Delta-c_p> 
{2\omega^2\over M_\Delta^2-M_p^2} \simeq 1.7\times
10^{-25}\;[\omega/\omega_0]^2\;,} 
reaction \eform\ is 
forbidden for all $E$.  Recalling that the photons are  thermal, we
see that if $c_\Delta-c_p\sim   \hat{\delta}(\omega_0)$, 
the GZK
cutoff due to resonant $\Delta(1232)$ formation would be relaxed.
Should it much exceed this value, formation would be precluded 
off virtually all CBR photons.

Reaction \eform\ is the dominant process leading to the GZK
cutoff.
If it is forbidden,
a  weakened version of the cutoff
may result from non-resonant photo-production: 
\eqn\eprod{p+ \gamma\ {\rm (CBR)}\longrightarrow p + \pi\,.}
If  $c_\pi=c_p$,
the threshold energy is
$E_p=M_\pi(2M_p+M_\pi)/4\omega$.  If 
$c_\pi-c_p>0$ 
the  threshold is larger.
As $E\rightarrow \infty$, the pion energy $E_\pi$ 
must remain finite.  Energy conservation yields the kinematic condition:
$$ 2\omega \ge  (c_\pi-c_p)\,E_\pi + {m_\pi^2\over 2E_\pi}\,,$$
which may be satisfied iff: 
\eqn\eproda{ c_\pi-c_p < \tilde\delta(\omega)\equiv 
{2\omega^2\over m_\pi^2}\simeq
5\times 10^{-24}\ [\omega/\omega_0]^2\,. }
For $c_\pi-c_p>\tilde\delta(\omega)$,
reaction \eprod, as well as multiple pion production, is
kinematically forbidden
off photons of energy $\omega\;$ 
 {\it at all proton energies.} 
For the actual thermal photons, 
$c_\pi-c_p\sim \tilde \delta(\omega_0)$ would suppress 
photo-pion production, or even eliminate it entirely 
so that no vestige of the GZK cutoff survives. 
Much larger, and experimentally intolerable, violations of
Lorentz invariance would be needed to affect
the interactions of UHE
cosmic rays with nuclei in the atmosphere.

A tiny value of the 
Lorentz-violating
$c_n-c_p$ stabilizes UHE neutrons. 
Tiny values of the parameters 
 $c_\Delta-c_p$ or $c_\pi-c_p$  forbid 
the processes underlying the GZK cutoff. Let's go for broke, and suppose
both 
Lorentz-violating effects are present. Then
cosmic-ray events of the highest energies could be
produced by primary UHE {\it neutrons\/}  from sources at cosmological
distances. They can have been 
stabilized and made GZK-resistant by 
departures from strict Lorentz invariance. They are electrically
neutral, so that they are undeflected
by magnetic fields  and can reveal their distant origins.
\medskip

Existing tests of special relativity are far too weak
to exclude these dramatic effects on UHE cosmic rays.
(See ref. \rcg\ for a list of current bounds.) Fortunately,
some bounds can be  strengthened considerably.
Laboratory tests of Lorentz invariance far
more precise than any done before are now feasible~\ref\rfort{E.N. Fortson,
private communication.}. Dedicated searches for velocity oscillations of
solar neutrinos, or of accelerator-produced $\sim\,$TeV neutrinos at
baselines of $\sim\! 1000$~km,  can reveal neutrino velocity differences as
small as $10^{-25}$. 
\medskip

\listrefs
\bye